\documentclass[prl,longbibliography,twocolumn,include graphics,final]{revtex4-2}

\usepackage[utf8]{inputenc}
\usepackage{amssymb}
\usepackage{amsmath}
\usepackage{amsfonts}
\usepackage{enumitem}

\usepackage[usenames,dvipsnames]{xcolor}
\usepackage[pdftex]{graphicx}
\usepackage{bm}

\usepackage{nccmath} % To resize the matrices (creating new environments).
\usepackage{array}
\usepackage{hyperref}
\usepackage{pifont}

\hypersetup{plainpages=false,colorlinks=true,linkcolor=Red, citecolor=blue, urlcolor=blue}
\newcolumntype{L}[1]{>{\raggedright\let\newline\\\arraybackslash\hspace{0pt}}m{#1}}
\newcolumntype{C}[1]{>{\centering\let\newline\\\arraybackslash\hspace{0pt}}m{#1}}
\newcolumntype{R}[1]{>{\raggedleft\let\newline\\\arraybackslash\hspace{0pt}}m{#1}}

\usepackage{changes}

\begin{document}

% Title layout

%\title{A~pseudogap~outside~the~usual~antiferromagnetic hot-spots~in~the~half-filled,~2D~triangular~lattice}
\title{Doping the Mott insulating state of the triangular-lattice Hubbard model reveals the Sordi transition}
%\title{\replaced{Doping the}{Filling-induced} Mott \added insulating state of \deleted{transition and pseudogap physics in } the triangular-lattice Hubbard model}
%\title{The pseudogap in the half-filled, 2D triangular lattice.}
\author{P.-O. \surname{Downey}$^1$}
\author{O. \surname{Gingras}$^{2}$}
\author{C.-D. \surname{Hébert}$^1$}
\author{M. \surname{Charlebois}$^3$}
\author{A.-M. S. \surname{Tremblay}$^1$}

\affiliation{$^1$D\'epartement de physique and Institut quantique, Universit\'e de Sherbrooke, Qu\'ebec, Canada J1K 2R1}
\affiliation{$^2$Center for Computational Quantum Physics, Flatiron Institute, 162 Fifth Avenue, New York, New York 10010, USA}
\affiliation{$^3$D\'epartement de Chimie, Biochimie et Physique, Institut de Recherche sur l’Hydrog\`ene, Universit\'e du Qu\'ebec \`a Trois-Rivi\`eres, Trois-Rivi\`eres, Qu\'ebec G9A 5H7, Canada}

\date{\today}


%  Abstract

\begin{abstract}
It has been reported that upon doping a Mott insulator, there can be a crossover to a \replaced{pseudogaped}{strongly correlated} metallic phase followed by a first-order transition to another thermodynamically stable metallic phase.
We call this first-order metal-metal transition the Sordi transition. 
\replaced{% While the earliest work on the Sordi transition using cluster dynamical mean field theory on $2\times2$ plaquettes had suggested that the transition should exist at finite temperature, bigger clusters on recent work have suggested that the transition may not be observable in real bi-dimensional lattices, i.e. that it exists at finite temperature. In this paper, we use a triangular lattice to argue that once long-range antiferromagnetic fluctuations are frustrated, the Sordi transition exists at finite temperature. Furthermore, using dynamical cluster approximation, we show that the transition is not an artifact of the cellular dynamical mean-field theory, suggesting that the transition is universal.
It was argued that the initial reports of Sordi transitions at finite temperature could be explained by finite size effects and biases related to the model and method used.
In this work, we report the Sordi transition on larger clusters at finite temperature on a triangular lattice, where long-range antiferromagnetic fluctuations are frustrated, using a different method, the dynamical cluster approximation instead of the cellular dynamical mean-field theory. 
This demonstrates that this first-order transition is a directly observable transition in doped Mott insulators and that it is relevant for experiments on candidate spin-liquid organic materials. 
}{To show theoretically that this transition is observable, it is important to provide calculations in situations where magnetic phase transitions do not hide the Sordi transition. 
It is also important to show that it can be found on large clusters and with different approaches. Here, we use the dynamical cluster approximation to reveal the Sordi transition on a triangular lattice at finite temperature in situations where there is no long-range magnetic correlations. }
\deleted{This is relevant for experiments on candidate spin-liquid organics. We also show that the metallic phase closest to the insulator is a distinct pseudogap phase that occurs because of strong interactions and short-range correlations.}
% -Abstract :\\
% \begin{itemize}
%     \item Sordi transition at finite temperature in a 2D lattice ;
%     \item DCA, triangular lattice, no long range AFM fluctuations
%     \item larger cluster
%     \item strong coupling pseudogap in a triangular lattice
% \end{itemize}
\end{abstract}

\maketitle

%%%%%%%%%%%%%%%%%%%%%%%%%%%%%%%%%%%%%%%%%%%%%%%% Introduction %%%%%%%%%%%%%%%%%%%%%%%%%%%%%%%%%%%%%%%%%%%%%%%%%%%%%%%%%%%%%%%%%%

%%

% Parler moins longtemps du MI et MT

Emergent states of strongly correlated electronic systems are among the most \replaced{fascinating }{interesting} phenomena in physics.
Simple models can capture the essence of intriguing correlated states while \replaced{remaining }{being} interpretable.
The Hubbard model~\cite{hubbard1963electron, hubbard1964electron2, hubbard1964electron3, Gutzwiller:1963,kanamori1963electron}, \replaced{although }{while} extremely simple to write and credited with important successes, still has no exact solution in two dimensions (2D) and remains a source of important challenges. 

Hubbard already understood that this model should contain the physics of the so-called Mott transition, namely a transition from a metal to an insulator caused by strong electronic repulsion.
% While it was already predicted in 1949~\cite{Mott:1949}, the Mott transition (MT), a transition from a metal to a MI caused by strong electronic repulsion, still captures a lot of attention. 
By now, numerous experiments have suggested that there is such an interaction-driven transition in the half-filled single-band model and that it is first-order~\cite{mcwhan_metal-insulator_1973,granados1993,Lefebvre:2000,dumm:2009, kanoda_mott_2011,Pustogow_Bories_2018, gorelovNatureMottTransition2010, alexanderDestructionMottInsulating1999}, a result supported by various embedding methods such as dynamical mean field theory (DMFT)~\cite{georges_hubbard_1992,Jarrell:1992,georges:1996, LTP:2006, vucicevic:2013} and its multi-site extensions: cluster DMFT (CDMFT)~\cite{park:2008, Balzer_Kyung_Senechal_Tremblay_Potthoff_2009, walshThermodynamicInformation:2019} and \added{the} dynamical cluster approximation (DCA)~\cite{Dang:2015,downeyMottTransitionWidom2023, LeeDualFermions:2008}.

Here we focus on the doping-driven Mott transition. 
Early on, several works have suggested that, upon changing the chemical potential $\mu$ away from half-filling, the \replaced{electronic compressibility, $\partial n/\partial \mu$ with $n$ the average number of electron per site, }{chemical potential dependence of doping} displays hysteresis characteristic of a first-order transition~\cite{Kotliar:2002, Macridin:2006, GullMomentum:2009, Sordi:2010, Sordi:2011}. 
There is, however, disagreement on the nature of the phases that are linked by doping-driven transitions. 
Single-site DMFT~\cite{Kotliar:2002} finds a first-order transition between the Mott insulator (MI) and the metal, while DCA cluster calculations find momentum space differentiation~\cite{GullFerrero:2010} or a  first-order transition on the electron-doped side only when the model is frustrated~\cite{Macridin:2006,GullMomentum:2009,Khatami:2010}. 
In contrast, Ref.~\citenum{Sordi:2010} found that a first-order doping-driven transitions could connect two thermodynamically stable metallic phases at finite temperature~\footnote{By thermodynamically stable, we mean a phase whose free-energy is a global minimum, contrary to metastable phases whose free energy are local minima, thus allowing phase coexistence.}: a pseudogap (PG) phase and a correlated Fermi liquid (cFL).
We refer to this type of doping-driven transition as the {\it Sordi transition}.

This transition could have profound implications as it appears to be the hallmark of the onset of a strongly correlated PG~\cite{Hankevych:2006,Sordi:2012}, namely a pseudogap that is a consequence only of short-range spin correlations~\cite{Senechal:2004,kyung:2006b}, not of long-wavelength spin fluctuations~\cite{Vilk:1997}. 
\replaced{
%What is meant by pseudogap here and in cuprates refers to the development of a momentum-dependent loss of spectral weight at the Fermi level, signaled in this work by the appearance of a local minimum. This will be explained and shown in more detail in Fig.~\ref{fig:phases}.
What is meant by pseudogap in this work simply refers to the appearance of a local minimum in the density of states, exemplified in Fig.~\ref{fig:sordi_transition}~(b) and quantified in the Supplemental Material~\cite{supplemental}. Notice that the PG phase is different from the bad-metal or bad-insulator phases discussed in single-site DMFT~\cite{Georges:2013,vucicevic:2013} since in these high-temperature phases one does not observe sharp peaks surrounding the minimum at the Fermi level.}{What is meant by pseudogap here and in cuprates will become clear later.}
In short, it is a phase where density of states and spin susceptibility are reduced and where various features appear in transport. 
Its onset for decreasing temperature happens at $T^*(p)$ that depends on doping $p$. 
%In particular, the doping at which the PG stops, $p^*$, is precisely where the Sordi transition occurs, implying that the PG might be related to Mott physics itself, affecting our theoretical understanding of the PG. 
The Sordi transition \added{also} explains several features of the pseudogap in cuprates~\cite{Sordi:2012,SordiResistivity:2013,SordiSuperconductivityPseudogap:2012,Reymbaut:2019,Walsh_Sound_2022,Sordi_Specific_heat_2019,Walsh_Sordi_Opalescence_2019, fournierTwoLinearScattering2024}, for example the experimentally observed sudden drop of $T^*$ at a critical doping $p^*$~\cite{doiron-leyraudPseudogapPhaseCuprate2017,CyrChoinieres_Nernst:2017}, critical opalescence~\cite{Walsh_Sordi_Opalescence_2019,Campi_Bianconi_2022}, the specific heat maximum~\cite{Sordi_Specific_heat_2019,Michon_Specific_Heat_2019} plus logarithmic temperature dependence~\cite{Reymbaut:2019,Michon_Specific_Heat_2019}, and even possibly high temperature linear scattering rate~\cite{fournierTwoLinearScattering2024,grissonnanche_linear-temperature_2021}.

More generally, the Sordi transition can be defined as a first-order transition between a strongly correlated metal and a weakly correlated metal. 
It is important to contrast the first-order nature of such a transition to general doping-driven transitions~\cite{Kotliar:2002, GullFerrero:2010, Sordi:2011, fayePseudogaptometalTransitionAnisotropic2017, bragancaCorrelationDrivenLifshitzTransition2018, fratinoDopingdrivenResistiveCollapse2022, chatzieleftheriouMottQuantumCritical2023, KotliarRMP:2006, LTP:2006, Macridin:2006}.

The Sordi transition was \replaced{first reported}{found} in Ref.~\citenum{Sordi:2010} using CDMFT on the 2D square lattice single-band Hubbard model with only nearest neighbor hopping $t$. \added{In Ref.~\citenum{fratino2016organizing}, the transition has also been seen when including a next-nearest hopping term $t'=-0.1t$.}  
\replaced{Similar phenomena have been}{As a metal-metal transition, it has been} observed in different contexts
\replaced{
~\cite{fratinoDopingdrivenResistiveCollapse2022, chatzieleftheriouMottQuantumCritical2023, bragancaCorrelationDrivenLifshitzTransition2018, FratinoChargeTransfer:2016}.
}{
Ref.~\citenum{fratinoDopingdrivenResistiveCollapse2022} observes it in the infinitely connected dimer Hubbard model at finite temperature, while Ref.~\citenum{chatzieleftheriouMottQuantumCritical2023} sees this transition in a multi-orbital Hubbard model on the Bethe lattice at finite temperature. 
In Refs.~\citenum{bragancaCorrelationDrivenLifshitzTransition2018}, it is detected in the one-band Hubbard model on the 2D square lattice at zero temperature and in Ref.~\citenum{FratinoChargeTransfer:2016} in the three-band model at finite temperature.
}

Objections that have been raised to the physical relevance of \replaced{the Sordi }{this} transition include the fact that in real systems, this transition could be hidden by long-range ordered phases or by long-range correlations precursors of zero-temperature ordered phases.
The latter argument has also been presented against the interaction-driven Mott transition~\cite{Schaefer_Fate:2015}. 
It has also been argued that the Sordi transition could be an artifact of CDMFT on a small 2$\times$2 square cluster\replaced{, as it }{that} overemphasizes singlet correlations.  
%Furthermore, it has been argued that because of the Mermin-Wagner theorem, such a Sordi transition should not be found on a 2D lattice at finite temperature; but all these works are either at zero $T$ or at dimensions larger than two.
%However according to the same Mermin-Wagner argument, if the Sordi transition is caused by an underlying long-range order, it shouldn't be observed at finite temperature in the one-band Hubbard model on a 2D lattice.

Here we study the Sordi transition on the triangular lattice. \added{While several magnetic phases appear at zero temperature for values of interaction slightly larger than the Mott transition~\cite{Schultz_Khoury_Desrochers_Tavakol_Zhang_Kim_2024},} \replaced{the triangular lattice}{that} geometrically forbids long-range magnetic correlations until very low temperature, in particular for the range of interaction strengths of interest to us: $8.5 \leq U/t \leq 10.54$~\cite{wietek_mott_2021,morita_nonmagnetic_2002, kyung:2006, sahebsara_hubbard_2008, laubachPhaseDiagramHubbard2015a, Misumi_Mott_triangular:2017, Tocchio_backflow_Mott:2008, Yoshioka_triangular:2009, yang_effective_2010, szasz_chiral_2020, chenQuantumSpinLiquid2022}.  
We also find that the strong coupling metallic phase is a PG phase in the triangular lattice away from half-filling~\cite{supplemental}.

\added{In this letter,} we achieve two broad goals.
\vspace{0cm}
\begin{enumerate}[label=(\roman*)]
    \item First, we show that the Sordi transition is not an artifact of either a) the square lattice, since we work on the triangular lattice, b) nor of CDMFT, since we use DCA, c) nor of four-site clusters, since we study six-site clusters, d) nor of long-correlation lengths, since magnetism is frustrated on the triangular lattice; 
We instead confirm that the transition is related to Mott physics, namely strong interactions, and that it should be observable\added{ at $T>0$}. 
    \item Second, we map out the phase diagram as a function of interaction $U$ and chemical potential $\mu$\added{ for two different temperatures} and identify \added{possibly observable} pseudogap regions\deleted{that might be observable}, in particular in spin-liquid candidate $\kappa$-BEDT layered organic compounds~\cite{oike:2017, yoshimoto:2013,yamamoto:2013}. 
\end{enumerate}

\paragraph*{The model. ---} 
We investigate the one-band 2D Hubbard model~\cite{Arovas_Berg_Kivelson_Raghu_2022,qinHubbardModelComputational2022a,LeBlanc_2015,Schaefer_Wentzell_2021}, defined by the Hamiltonian
\begin{align}
  \label{eq:H}
    H= & 
      -\sum_{i, j,\sigma} t_{ij} \hat{c}^\dagger_{i\sigma} \hat{c}_{j\sigma} 
      + U \sum_i \hat{n}_{i\uparrow} \hat{n}_{i\downarrow}
      - \mu \sum_{i\sigma} \hat{n}_{i\sigma},
\end{align} 
where $t_{ij}$ represents the hopping terms between \deleted{near-neighbor} sites $i$ and $j$, $\hat{c}^\dagger_{i\sigma}$ and $\hat{c}_{i\sigma}$ are respectively the creation and annihilation operators for electrons with spin $\sigma$ at site $i$, $U$ is the strength of the on-site Coulomb interaction between electrons, $\hat{n}_{i\sigma} = \hat{c}^\dag_{i\sigma}\hat{c}_{i\sigma}$ is the number operator for electrons with spin $\sigma$ at site $i$, and $\mu$ is the chemical potential.
The six-site cluster that we study is shown in Fig.~\ref{fig:sordi_transition}~(a). We consider only the nearest-neighbor hopping terms $t$ and $t'$, where $t'$ is useful to model the anisotropic triangular lattice and looks like a second-neighbor hopping when the lattice is mapped on a set of orthogonal basis vectors. 
Here we take $t=-t'=1$. 
These signs of the hopping terms were chosen with the cuprate convention. 
At half-filling, the sign of $t'$ is irrelevant and, in general, changing it is equivalent to a particle-hole transformation~\footnote{This is explained in Appendix~A of Ref.~\cite{downeyMottTransitionWidom2023} based on the anisotropic triangular lattice.}. 
We work in units where hopping $t$, lattice spacing, Planck and Boltzmann constants are unity.
\deleted{The temperature in every figure \added{where it is not specified} is $T=1/15$. }

%\added{Note that with four-site CDMFT, the Sordi transition in addition to the square lattice with $t'=0$, the transition has also has also been seen for $t'=-0.1t$~\cite{fratino2016organizing}. for the three-band Hubbard model~\cite{FratinoChargeTransfer:2016} and on the anisotropic triangular lattice ($t'=0.4t$) ~\cite{hebert_superconducting_2015} . In all these cases, the possible role of antiferromagnetism can be invoked. Here, we avoid this possible interference}. 

The solutions to this model are obtained using a state-of-the-art DCA implementation~\cite{hettler_dynamical_2000, maier_quantum_2005,Pavarini_Koch_Coleman:2015} with continuous-time auxiliary-field quantum Monte Carlo interaction expansion (CT-AUX)~\cite{Gull_continuous_2008, Gull:2011}.
%Both CDMFT and DCA are extension of DMFT, but DCA works with the reciprocal lattice space and is constrained to translation invariant solution. 
The six-site cluster on Fig.~\ref{fig:sordi_transition} was shown in Ref.~\citenum{downeyMottTransitionWidom2023,fournierTwoLinearScattering2024} to exhibit the major features seen in its larger counterpart (12 sites). 
Note that, contrarily to CDMFT, the DCA cluster has periodic boundary conditions  \replaced{on the cluster}{and translation-invariant solutions}.

\begin{figure}
    \centering
    \includegraphics[width=\linewidth]{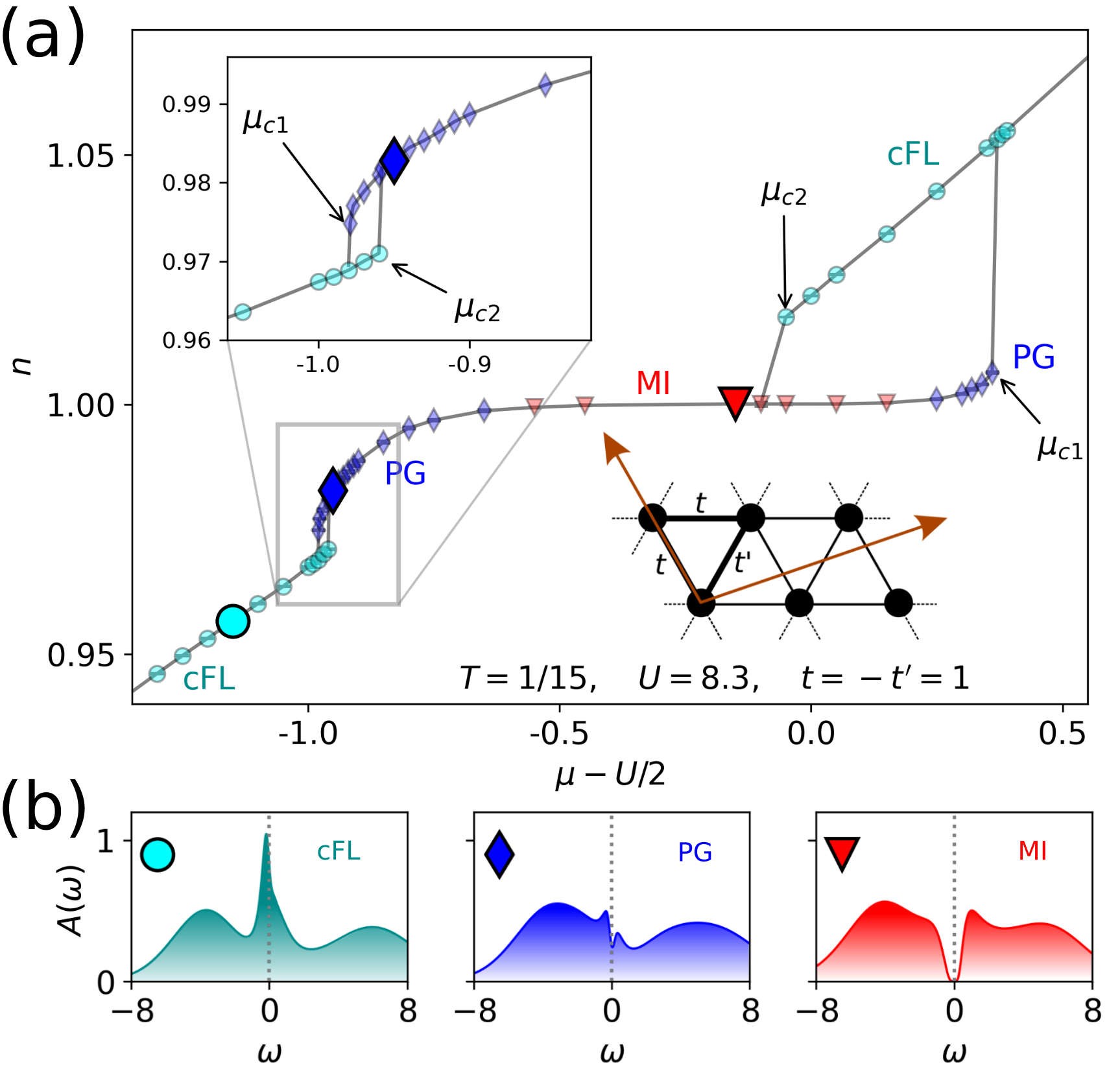}
    \caption{
    \added{
    (a) First-order doping-driven Sordi transitions at $U=8.3$ and $T=1/15$.
    The hole-doped Sordi transition is highlighted in the inset.
    We define the quantities $\mu_{c1}$ and $\mu_{c2}$ as the critical $\mu$ for the transition from the Mott insulator (MI) or pseudogap (PG) to correlated Fermi liquid (cFL), and from cFL to MI or PG, respectively.
    % \added{These phases are explained and shown in Fig.~\ref{fig:phases}.}
    %The Sordi transition corresponds to a transition where $\mu_{c2}$ connects to a state that is not at half-filling.
    The Sordi transition corresponds to a first-order transition that evolves from the Mott transition at half-filling but that occurs between two stable metallic states.
    The cluster used on the real lattice is displayed in the lower right corner.
    (b) Local spectral weight for the representative points of (a). We use the spectral weight to classify the states obtained by the calculations. More details can be found in the Supplemental Material~\cite{supplemental}.
    }
    }
    \label{fig:sordi_transition}
\end{figure}

\paragraph*{Filling induced first-order transitions. --- }
Figure~\ref{fig:sordi_transition} presents the first-order doping-driven transitions obtained on both the hole- and electron-doped sides, for $U=8.3$\added{ and $T=1/15$}.
The hole-doped doping-driven transition highlighted in the inset is clearly a Sordi transition\added{ since it shows a first-order transition between two metallic (compressible) states away from half-filling}. 
As in the careful studies of Refs~\citenum{Sordi:2010} and \citenum{Sordi:2011} on the square lattice, \replaced{observing this transition requires very fine scans in $\mu$ and }{it needs very fine scans in $\mu$ to be observed and it} might have been missed in earlier studies~\cite{Macridin:2006, GullFerrero:2010}.
One needs to start from a previously converged solution in the MI and to make small steps in $\mu$ in order to numerically reveal the hysteresis.

%although we see the onset of a PG phase, it is most likely metastable, since it happens deep in the hysteresis.
%Hence, we do not label the DDT on the electron doped side as a Sordi transition: the first-order transition for this set of parameters is between a cFL and a MI.
%As in Ref.~\citenum{Sordi:2011}, a MI is found at half-filling and a PG on both the hole and electron doped sides at low doping.
%Increasing doping further, the PG phases ends at a first-order transition to the cFL states.

\begin{figure}
    \centering
    \includegraphics[width=\linewidth]{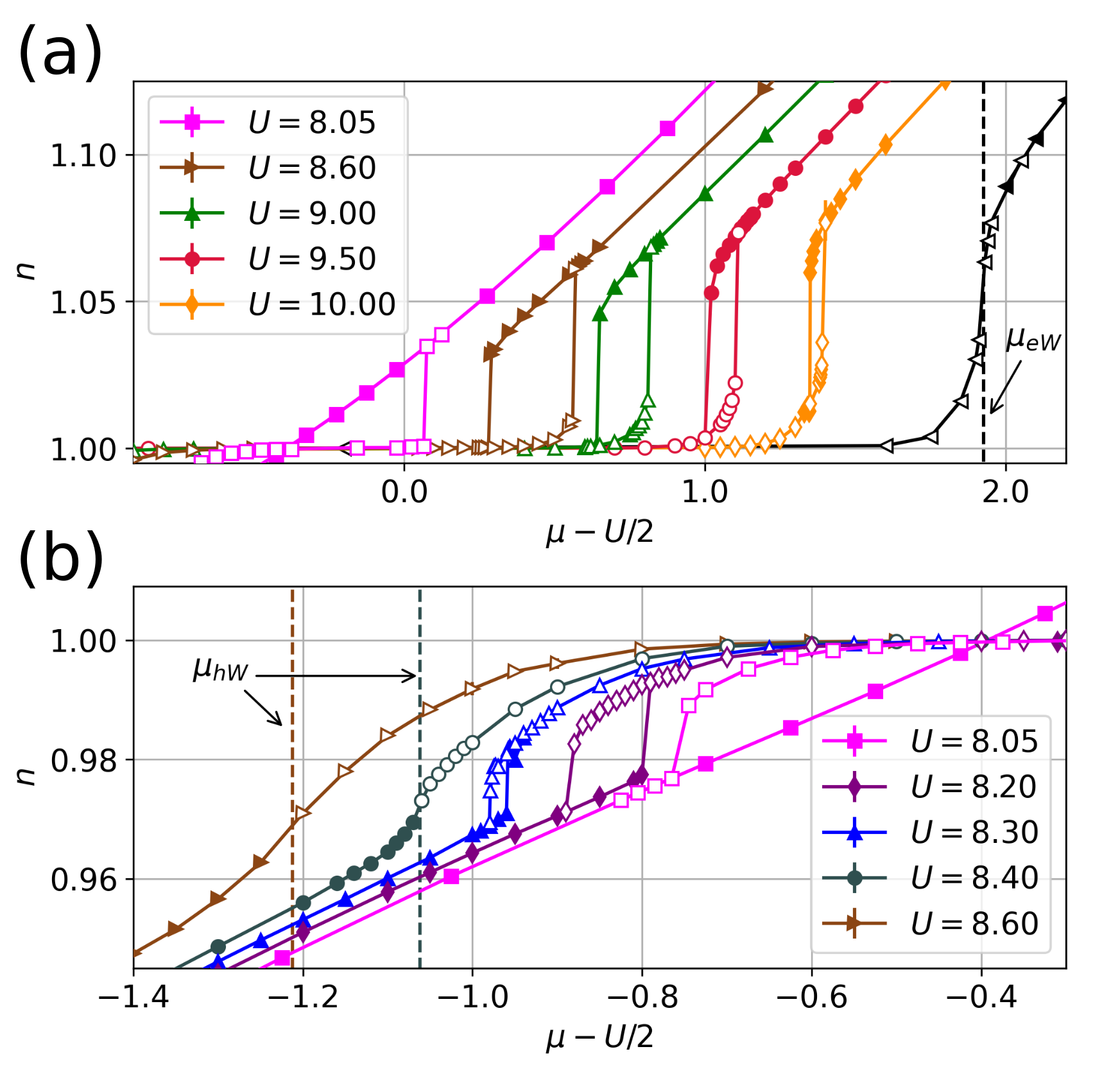}
    \caption{Evolution of the first-order doping-driven transition with respect to $U$ on the (a)~electron- and (b)~hole-doped sides at $T=1/15$.
    The Widom line $\mu_{hW}$ ($\mu_{eW}$) is the value of $\mu$ on the hole-doped (electron-doped) side where the compressibility $\partial n /\partial \mu$ is maximal \added{for a given value of $U$}.}
    \label{fig:sordi_doped}
\end{figure}

% changed the text from a paragraph to the other
As expected, on a triangular lattice, the hole- and electron-doped sides are asymmetrical. \replaced{For these parameters on}{On} the electron-doped side, the hysteresis is between a metal and an insulator, as observed on the square lattice with second-neighbor hopping~\cite{Macridin:2006, Khatami:2010}. 
Nevertheless, what looks like an hysteresis between a metal and an insulator in fact evolves continuously into the Sordi transition, as shown in Fig.~\ref{fig:sordi_doped}~(a) \replaced{that}{.
It} displays the evolution of the doping-driven transitions as a function of interaction strength $U$. 
Even if a first-order metal to insulator transition seems to be more commonly witnessed~\cite{Macridin:2006,Khatami:2010}, increasing $U$ makes it clear that there is a Sordi transition.
The Sordi transition was often presented as a transition induced by hole doping~\cite{Sordi:2012, chatzieleftheriouMottQuantumCritical2023, bragancaCorrelationDrivenLifshitzTransition2018,fratinoDopingdrivenResistiveCollapse2022}, but our results suggest that the transitions on the electron- and hole-doped sides are very similar, even in the case of large electron-hole asymmetry.

Figure~\ref{fig:sordi_doped}~(b) presents the hole-doped side\replaced{. We }{where we} see that the Sordi transition disappears upon increasing $U$ \replaced{to }{between} $U=8.35 \pm 0.05$ \replaced{and that it then becomes }{, where it is connected to} a Widom line $\mu_{hW}(U)$~\cite{Sordi:2012}, \added{that is} a line of maxima of thermodynamic quantities~\cite{XuStanleyWidom:2005,Simeoni:2010}\replaced{. In this case, the Widom line is defined as a maximum of the }{, here} compressibility $\partial n/\partial \mu$, that slowly vanishes as $U$ increases.
\replaced{On the other hand, for $U$ down }{Decreasing $U$} to $U=8.05$, once in the metallic state \replaced{it cannot }{, there is no way to} go back to the insulator.
This is the expected behavior when $U$ is in the interval $U_{c1}< U< U_{c2}$, the spinodals of the interaction-driven Mott transition at half-filling. 
Note that the Sordi transition on the electron-doped side exists at that temperature for a much wider range of interaction ($8.8\lesssim U\lesssim 10.3$) than on the hole-doped side.

% \begin{figure}
%     \centering
%     \includegraphics[width=\linewidth]{figures/3-electron_doped.png}
%     \caption{ }
%     \label{fig:electron_doped}
% \end{figure}

%The evolution of the DDT with the interaction strength $U$ on the electron doped side presented in Fig.~\ref{fig:sordi_transition}~b) reveals a slightly different behavior.

\begin{figure}
    \centering
    \includegraphics[width=\linewidth]{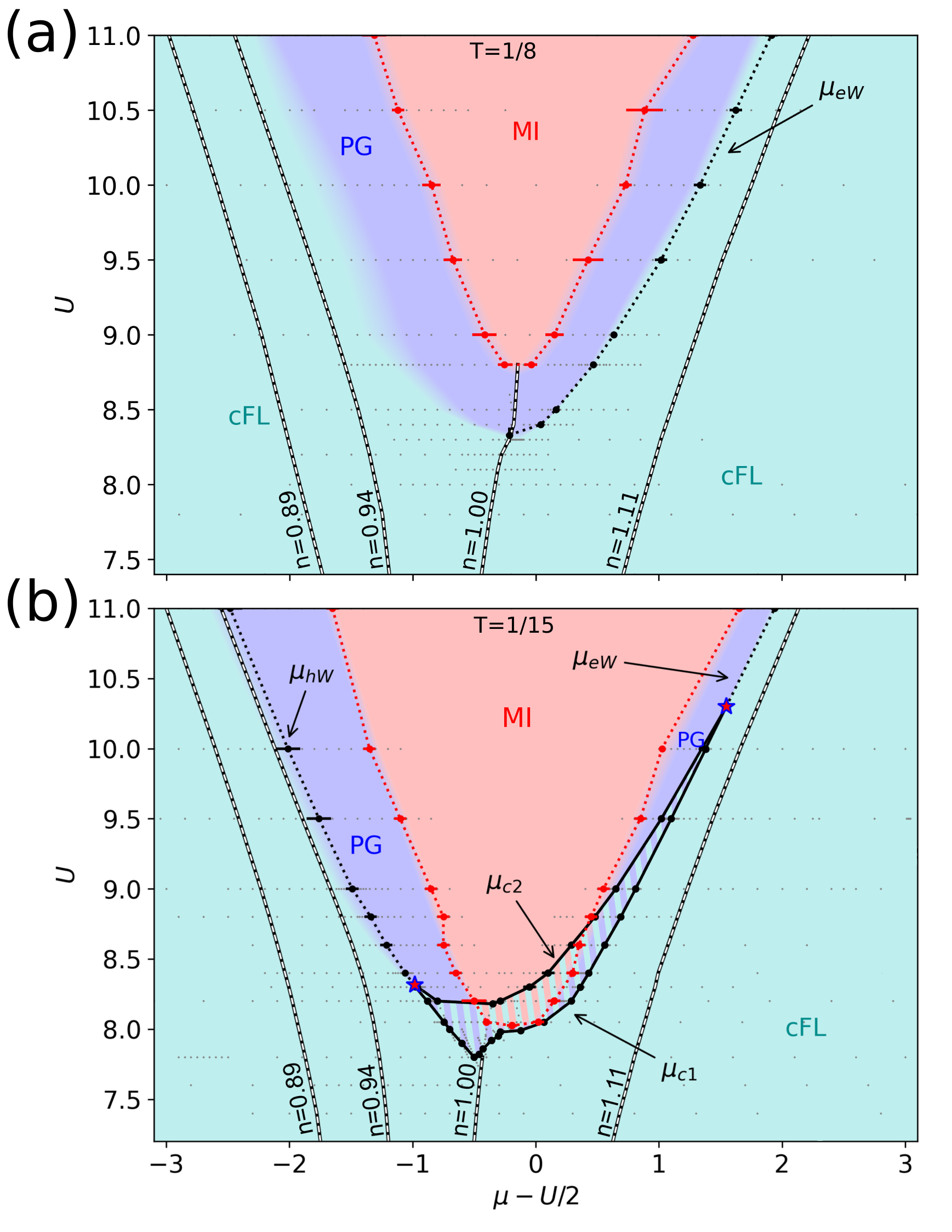}
    \caption{
    Phase diagram with $U$ on the vertical axis and $\mu-U/2$ on the horizontal axis for (a) $T=1/8$ and (b) $T=1/15$ showing the connections between the MI, Widom lines $\mu_{eW}$ and $\mu_{hW}$, and Sordi transitions.
    \replaced{Phases are labelled as }{It also shows the regions where there is a} correlated Fermi liquid (cFL), \deleted{a} pseudogap (PG) and \deleted{a} Mott insulator (MI).
    The red \replaced{stars with blue contours}{and blue stars} highlight the critical $(U,\mu)$ of the Sordi transition where the compressibility tends toward infinity, on both the hole- and electron-doped sides.
    % It is extracted by extrapolating the two nearest $U$ for which there was a first-order transition (give the values of the stars).
    The color gradients between some of the phases represent the crossover nature of the transitions. \added{The red dotted line represents where the spectral weight at $\omega=0$ is below a small threshold ($\delta=0.005$)~\cite{supplemental}.}
    }
    \label{fig:mu_U}
\end{figure}

\paragraph*{Connection to Mott physics. ---} To better grasp the connection between the Sordi transition, the PG and Mott physics, we present their interplay on the phase diagrams presented in Fig.~\ref{fig:mu_U}.
It displays the three phases, MI, PG and cFL, encountered in Fig.~\ref{fig:sordi_doped} for a wide range of $\mu$ and $U$ \replaced{for two different temperatures: $T=1/8$ in (a) and $T=1/15$ in (b). Notably, the doping range of the PG decreases with temperature, and the Mott energy gap becomes smaller as the temperature increases.
At $T=1/8$, the phase diagram is rather simple. We find that the Mott gap increases with $U$, and that a rapid crossover separates the electron-doped PG and the cFL. 
The physics becomes more interesting when the temperature decreases. At $T=1/15$, we find that the cFL can coexist with either the PG or the MI, as highlighted by the hatched region. Although we cannot reach infinitesimally small temperatures with continuous-time quantum Monte Carlo, we expect that the Sordi transition likely remains down to $T=0$ if long-range order does not appear~\cite{chatzieleftheriouMottQuantumCritical2023}. }{It also reveals the phase coexistence for the first-order Sordi transition, as highlighted by the hatched region.}
The portions of this region corresponding to Sordi transitions are those where the hysteresis occurs between the cFL and the PG phases at constant $U$.
The limit between the PG and the cFL are intimately related to the first-order doping-driven transition or to the Widom line.

\replaced{While the limit between the PG and the cFL is clear when there is a first-order transition, it is no longer the case at large $U$ or high temperature. Indeed, in these cases, the first-order transition evolves into a Widom line or leaves no visible trace in the compressibility $\partial n/\partial \mu$. It makes it particularly hard to define only one objective criterion for the boundary between the PG and the cFL~\cite{supplemental}. }{Although it is clear at larger $U$ that the PG is adiabatically connected to the cFL through a Widom line, defining the boundary is hard, as discussed below.}
Another important remark is that the spinodals $\mu_{c1}$ and $\mu_{c2}$ of the Sordi transition are connected at half-filling to $U_{c1}$ and $U_{c2}$\added{ discussed in Ref.~\citenum{downeyMottTransitionWidom2023}, showing the direct relationship between the Sordi transition and the Mott transition}. 
Finally, we observe around $\mu=3.4$ and $U=7.8$ that a PG can be found with a $U$ slightly lower than $U_c$ for the MI.
This peninsula in the PG phase has an analog in the 1D Hubbard model~\cite{Giamarchi:1997}.
%. Although we cannot explain it, this feature seem very similar to Giamarchi's bump in the phase diagram of the 1D Hubbard model~\cite{Giamarchi:1997}.

% \label{sec:PG_proof}

%Showing without doubt that we do have a PG is not straightforward, and the proof usually needed in that case may never be enough. Nevertheless, 

% \og{La section suivante n'est plus vraiment utile. À adapter.}
\paragraph*{\deleted{The pseudogap. ---}} 
\replaced{}{Here we show that the metallic phase that crosses over smoothly from the MI is a PG phase. For $U$ in the vicinity of $U_{c1}$ and $U_{c2}$, that phase is separated from a cFL by the Sordi transition.
% Here we show some convincing elements that are needed in order to conclude that we have a PG.
Figure~\ref{fig:phases} is an example case for $U=8.3$ and $T=1/15$.
Figure~\ref{fig:phases}~a) is like Fig.~\ref{fig:sordi_transition}, but where the phases are labeled.
The blue diamonds denote a phase that is metallic, since the compressibility $\partial n/\partial \mu$ is finite, yet is continuously connected to the MI, denoted by red inverted triangles.
Figures~\ref{fig:phases}~(c-e), shows the spectral weight \added{$A(\omega)$} for every phase\added{, defined as $-2\text{Im}G(\omega) = A(\omega)$}.
Figure~\ref{fig:phases}~(d) clearly shows a drop of spectral weight at the Fermi level $\omega=0$, compared to the peak of the cFL in (c), and the gap of the MI in (e).
This incomplete gap opening in (d) is a strong indication of the existence of a PG~\cite{kyung:2006b}.
% Fig.~\ref{fig:phases}~(c) shows the most important evidence for the PG. Indeed, on the middle panel, we have a clear drop of spectral weight at the Fermi level compared to the cFL on the left-hand panel, but we still have spectral weight, in contrast to what is seen on the right-hand panel (MI). This is known as an usual proof of the presence of a PG~\cite{kyung:2006b}.
% \begin{figure}
%     \centering
%     \includegraphics[width=\linewidth]{figures/S1-phases.png}
%     \caption{
%     Three elements supporting the existence of a PG in the triangular Hubbard model at $U=8.3$ and $T=1/15$.
%     (a) Same as Fig.~\ref{fig:sordi_transition}, but the phases are labeled as follow:  cyan circles, blue diamonds and red triangles represent respectively cFL, PG and MI phases.
%     (b) The ratio of spectral weight~\cite{note_spectral} at $\omega=0$ between patches 1 and 3 of the inset. The inset also shows the Fermi surface at $U=0$ and $n=1$ as a guide for the location of the Fermi surface with respect to the patches outside the MI.
%     The markers are the same as is (a).
%     (c) Local spectral weight for the representative points of (a) and (b).
%     \og{Shift $\mu$ to $\mu - U/2$.}
%     }
%     \label{fig:phases}
% \end{figure}

However, this is not sufficient to unequivocally conclude to the presence of a PG: 
In addition, we provide indications of a $\textbf{K}$-dependent change in the spectral weight at $\omega=0$.
Figure~\ref{fig:phases}~(b) shows a large change in the ratio between the spectral weight in different momentum patches, in contrast with the cFL at higher doping where the ratio changes only slightly.
Since both the hole- and the electron-doped regions have large changes in this ratio, we conclude that both exhibit PGs. 

Note that Fig.~\ref{fig:phases}~(b) suggests that the mechanism of the PG does not involve 120$^\circ$ antiferromagnetic (AFM) fluctuations, unlike Ref.~\citenum{downeyMottTransitionWidom2023}.
Indeed, as detailed in Ref.~\citenum{downeyMottTransitionWidom2023}, 120$^\circ$ AFM should involve equal loss of spectral weight in both patches 1 and 3, since both patches have only one 120$^\circ$ AFM hot spot.
Other types of very short-range magnetic correlations could be at play~\cite{wietek_mott_2021}, but it is not our ambition to tackle this subject in the present work.}

% \renewcommand\refname{References}

%Together with our most important result, the discovery of the Sordi transition, CDMFT is not the only technique where the latter transition exists at finite temperature. Plus, since we used a triangular lattice, there is no doubt that the Sordi transition does not originate from any magnetic transitions between the PG and the cFL. 

%We believe that Fig.~\ref{fig:mu_U} is our most intriguing figure as it shows how the Sordi transition is connected to the MT at half-filling. This is a clear proof that The Sordi transition is an essential part of Mott physics. Furthermore, since the PG stops at the Sordi transition, it seems very likely that the Mott physics is part of the mechanism involved in the formation of the PG.

%To prove experimentally that there is a PG and a Sordi transition, experimentalists face many important challenges. When using simulators~\cite{mingRealizationHoleDopedMott2017, tarruell_quantum_2018}, the temperature range is yet very limited. With layered organic materials, either the possible dopings in triangular systems are very limited~\cite{oike:2017, yoshimoto:2013}, or field effect transistors are needed~\cite{yamamoto:2013}. If one could overcome those difficulties, there is no way with layered organic materials to directly access to the Fermi surface and prove some $k$-dependence since ARPES isn't possible. Unfortunately, Widom lines such as $\mu_{hW}$ and $\mu_{pW}$ would be almost impossible to identify experimentally. A lot more theoretical work is thus needed to propose an experiment.

\paragraph*{Discussion. --- }
% Max 
% P-O 
%
Our results provide evidence for Sordi transitions in the triangular Hubbard model using DCA, hence showing that this phenomenon is not an artifact of two-orbital DMFT nor \replaced{ of CDMFT on $2\times2$ clusters. }{ CDMFT.
This observation of the Sordi transition in a six-site cluster shows that it is not a feature of only $2\times2$ clusters.} %Also, among the many clusters where the Sordi transition was found, we used the largest. It is relevant because an argument against the existence of the transition was the size of the cluster used in previous studies. 
We stress that our implementation of the CT-AUX impurity solver and DCA algorithm are independent of any previous studies on this subject.
Furthermore, many other studies found that the Hubbard model on a triangular lattice does not have long-range magnetic correlations at half-filling in the temperature and interaction regime we study ($\sim 8.2$)~\cite{morita_nonmagnetic_2002, kyung:2006, sahebsara_hubbard_2008, laubachPhaseDiagramHubbard2015a, Misumi_Mott_triangular:2017, Tocchio_backflow_Mott:2008, Yoshioka_triangular:2009, yang_effective_2010, szasz_chiral_2020, chenQuantumSpinLiquid2022, wietek_mott_2021}.
This suggests that our study is not biased by large correlation lengths, that the Sordi transition is not a magnetic transition and that it could be measured experimentally. %should be observable with the proper material. 

The Sordi transition separates two different metals: one highly correlated and the other less correlated.
\replaced{The former is interpreted as a PG state, although the formal proof of a $\textbf{k}$-dependent loss of spectral weight remains to be seen and is kept for future works, as it will require additional extensive calculations.}{We have shown that the former has a PG, meaning a $\textbf{k}$-dependent loss of spectral weight, as observed in other works
%~\cite{downeyMottTransitionWidom2023, wuEffectVanHove2020}.
}
Triangular lattices are highly frustrated, thus the existence of the strong-coupling PG in the triangular Hubbard model is strong evidence that long-range magnetic correlations are not a necessary ingredient for the onset of a PG, consistent with other studies~\cite{Senechal:2004,Hankevych:2006,kyung:2006b, Sordi:2011,Simkovic_Rossi_Georges_Ferrero_2022}.
%
%The mechanism in both references seems closely related to 120$^\circ$ AFM, but they are constrained at half-filling ($n=1$). However, here we are not at half-filling, and doping even slightly the material suppress this 120$^\circ$ AFM, as argued in more details in SM.

%but we show in SM that it cannot be the case. 
%Here, the mechanism seem different, as we show in SM that the PG cannot be linked to 120$^\circ$ AFM.
%
%The unset of this kind of PG can be understood the so-called Vilk criterion~\cite{Vilk:1995,Vilk:1997}, namely when the AFM correlation length exceeds the thermal De Broglie wavelength. Compared to the PG in Ref.~\citenum{downeyMottTransitionWidom2023}, the temperature is lower increasing the de Broglie wavelenght, and with doping, the magnetic correlation length lowers, making the Vilk criterion much less likely to apply. This suggests that the observed PG exists beyond the trivial weak-coupling mechanism, being instead in a strong-coupling mechanism. As supporting evidence, the SM shows that neither the hole-doped nor the electron-doped PG can be explained by 120$^\circ$ AFM fluctuations. 
%
%The existence of the strong-coupling PG in the triangular Hubbard model is a strong evidence that long-range magnetic correlations is not necessary ingredient for the onset of a PG. This conclusion is coherent other studies~\cite{kyung:2006b, Sordi:2011}.

As suggested by Fig.~\ref{fig:sordi_doped}, the hysteresis associated to the Sordi transition \replaced{appears to be indissociable from the Mott transition, as follows from the phase diagram, Fig.~\ref{fig:mu_U}~(b), where it clearly emerges from the Mott insulator. This was seen on the square lattice as well~{\cite{Sordi:2010,Sordi:2011}}.}{is a universal phenomenon, and from the phase diagram Fig.~\ref{fig:mu_U}~(b), it clearly emerges from the Mott insulator.}
At temperatures where DCA calculations can be performed, this hysteresis is found in the low interaction limit of the doped MI~\cite{Sordi:2010, Sordi:2011, Sordi:2012, SordiResistivity:2013}, but it should persist to larger interactions in lower temperature phase diagrams~\cite{Sordi:2010, Sordi:2011}. \added{The transition also disappears when the temperature increases.} 
\replaced{Furthermore, we find that at $T=1/15$, the critical $\mu_{c1}$ and $\mu_{c2}$ of the interaction-driven Mott transition are connected }{In Fig.~\ref{fig:mu_U}, you see that the Sordi transitions delimited by $\mu_{c1}$ and $\mu_{c2}$ on both doped sides are connected }to $U_{c1}$ and $U_{c2}$ \replaced{of the interaction-driven Mott transition}{from half-filling}. 
The PG is constrained within the boundary delimited by $\mu_{c1}$. This is further evidence that the PG is an unavoidable consequence of Mott physics in systems where singlet formation can occur because of strong interactions.

The scenario of a first-order doping-driven transition separating a strongly correlated metal from a weakly correlated one is similar to results for the two-orbital Hubbard model in Ref.~\citenum{chatzieleftheriouMottQuantumCritical2023} where the authors observed the Sordi transition between a strongly-correlated Hund metal and a weakly-correlated metal.
Their conclusion suggests the hypothesis that the two critical points denoted by stars in Fig.~\ref{fig:mu_U} are finite temperature critical points that might extend to quantum critical points at zero temperature.

Such fundamental connections between different phenomena have important consequences for our understanding of the Hubbard model and thus of the cuprates, modeled on a square lattice with \replaced{frustrated}{frustrating} hoppings.
Having a PG that is unrelated to long-range AFM correlations but that still starts at a critical endpoint suggests that the sudden drop of $T^*$ in cuprates near the critical doping, usually noted $p^*$, is a signature of Mott physics~\cite{SordiResistivity:2013,Reymbaut:2019,doiron-leyraudPseudogapPhaseCuprate2017,CyrChoinieres_Nernst:2017}.
Although the physics of the PG on the hole-doped side in our results is relevant for the PG found in hole-doped cuprates~\cite{Sordi:2011}, our PG on the electron-doped side is not related to the PG of the electron-doped cuprates, since those arise from long wavelength spin fluctuations~\cite{Hankevych:2006,gauvin-ndiayeResilientFermiLiquid2021, Senechal:2004, Weber:2010, horioOxideFermiLiquid2020, Motoyama:2007}.
In addition, our work is relevant for doped layered organic superconductors~\cite{oike:2017, yoshimoto:2013}, for field-effect doped organic superconductors~\cite{yamamoto:2013}, \added{for doped 1T-TaS$_2$~\cite{Law_Lee_2017}} or for analog simulators~\cite{mingRealizationHoleDopedMott2017, tarruell_quantum_2018,Yang_Liu_Mongkolkiattichai_Schauss_2021,Mongkolkiattichai_Liu_Garwood_Yang_Schauss_2022}. Our predictions motivate experiments at low temperatures \added{in these systems}.

\paragraph*{Conclusion. ---}
We report the Sordi transition between a strongly-correlated pseudogap phase and a correlated Fermi liquid on the triangular lattice Hubbard model at finite temperature.
\replaced{This transition is not an artifact of two-orbital DMFT, nor of $2\times 2$ CDMFT as confirmed by our \replaced{alternate $6$-site DCA approach}{translationally invariant solutions obtained using DCA}. We expect that other computational methods will be able to identify this transition on the triangular lattice, contrary to the unfrustrated square lattice where large system size calculations do not even find a Mott transition due to  antiferromagnetism~\cite{Schaefer_Fate:2015}.}{ transition is not an artifact of DMFT or CDMFT since we used translationally invariant solutions obtained using DCA}
On the triangular lattice, magnetic correlations arising from superexchange are short-ranged \added{at finite temperature}, so the mechanism for the pseudogap is not related to long-wavelength magnetic fluctuations.
Finally, our $U-\mu$ phase diagrams show that both this pseudogap and the Sordi transition are phenomena intimately associated to doping a Mott insulator, and as such they should be observable in any dopable system where the Mott transition can be observed at finite temperature.
Future work should focus on additional experimental predictions, in particular for layered organic materials. % investigate the phase diagram of the triangular lattice experimentally that could prove the existence of the strong-coupling pseudogap or the Sordi transition in layered organic systems.

%Those conclusions aren't new, but the fact that they can be found with another technique than CDMFT, here DCA, that the Sordi transition is not be a small cluster artifact, that the magnetic transition hypothesis is now extremely unlikely, and that, finally, the temperature at which the Sordi transition is found is much higher than in earlier work. Future theoretical work should continue to investigate the phase diagram of the triangular lattice and experimental work should verify the existence of the Widom line in layered organic systems.

\section*{Acknowlegments}
We thank Claude Bourbonnais, J\'er\^ome Fournier, Chlo\'e Gauvin-Ndiaye, Antoine Georges, Andrej Pustogow, Peter Rosenberg, Thomas Sch\"afer, David S\'en\'echal, Giovanni Sordi and Caitlin Walsh for fruitful discussions. We also thank Mo\"ise Rousseau for his technical support.
The computational resources for this work were provided by Compute Canada and Calcul Qu\'ebec. We acknowledge support by the Natural Sciences and Engineering Council of Canada through RGPIN-2019-05312 and by the Canada First Research Excellence Fund. 
The Flatiron Institute is a division of the Simons Foundation.

%\bibliography{references.bib}
  

%%%%%%%%%% Merge with supplemental materials %%%%%%%%%%

%%%%%%%%%% Prefix a "S" to all equations, figures, tables and reset the counter %%%%%%%%%%

\appendix

\pagebreak
~
\newpage

\setcounter{equation}{0}
\setcounter{figure}{0}
\setcounter{table}{0}
\setcounter{page}{1}

\renewcommand{\theequation}{S\arabic{equation}}
\renewcommand{\thefigure}{S\arabic{figure}}
% \renewcommand{\bibnumfmt}[1]{[S#1]}
% \renewcommand{\citenumfont}[1]{S#1}

%%%%%%%%%% Prefix a "S" to all equations, figures, tables and reset the counter %%%%%%%%%%

% \newpage
% \newpage

\onecolumngrid
\begin{center}

{\large\textbf{\boldmath
Supplemental Material\\ [0.5em] {\small to} \\ [0.5em]
Doping the Mott insulating state of the triangular-lattice Hubbard model reveals the Sordi transition}}\\[1.5em]

P.-O. \surname{Downey}$^1$, O. \surname{Gingras}$^{2}$, C.-D. \surname{Hébert}$^1$, M. \surname{Charlebois}$^3$, and A.-M. S. \surname{Tremblay}$^1$\\[0.5em]

\textit{\small
$^1$D\'epartement de physique and Institut quantique, Universit\'e de Sherbrooke, Qu\'ebec, Canada J1K 2R1\\
$^2$Center for Computational Quantum Physics, Flatiron Institute, 162 Fifth Avenue, New York, New York 10010, USA\\
$^3$D\'epartement de Chimie, Biochimie et Physique, Institut de Recherche sur l’Hydrog\`ene, Universit\'e du Qu\'ebec \`a Trois-Rivi\`eres, Trois-Rivi\`eres, Qu\'ebec G9A 5H7, Canada
}

\vspace{2em}
\end{center}

\twocolumngrid

\begin{figure}
    \centering
    \includegraphics[width=0.9\linewidth]{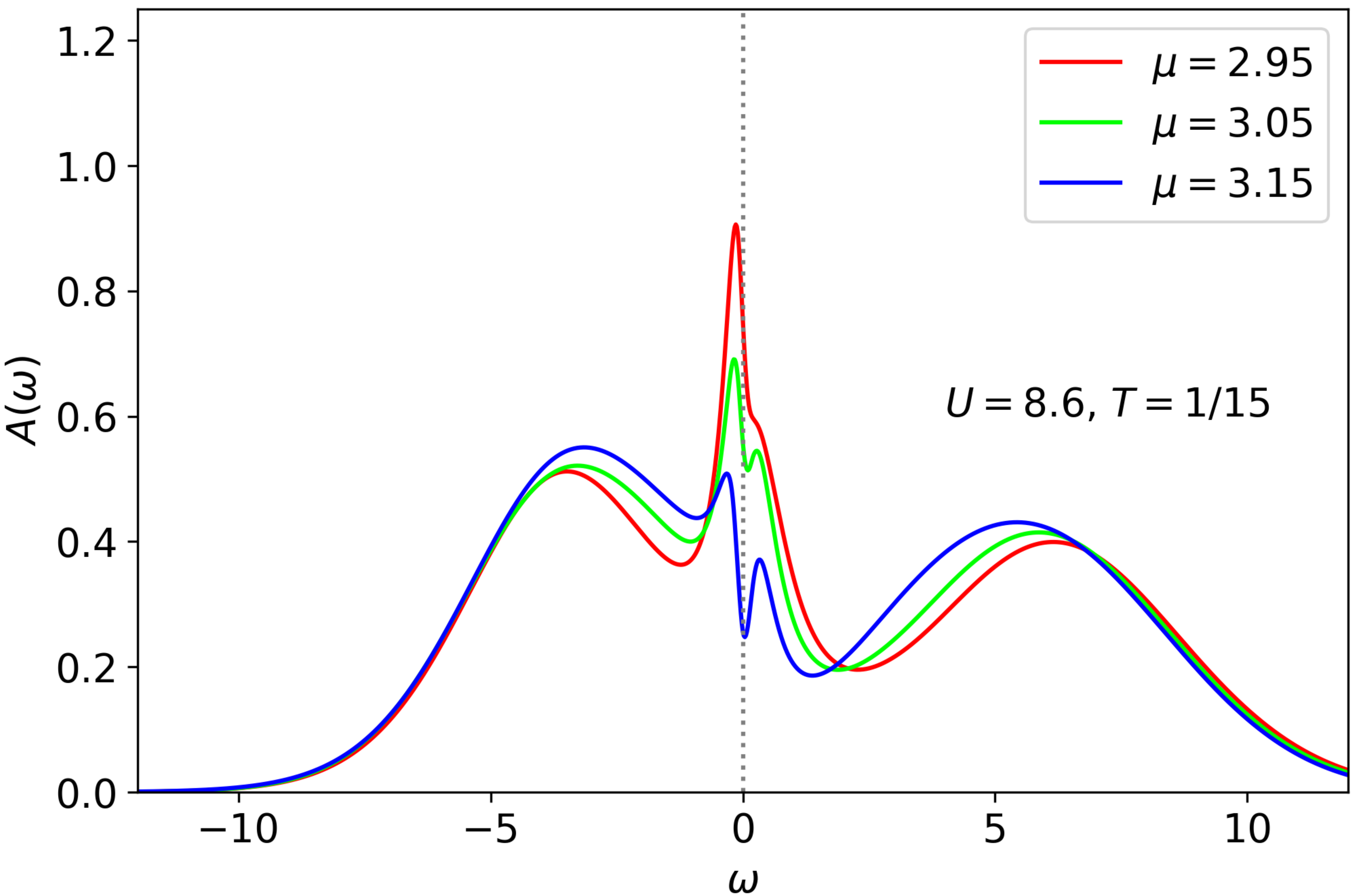}
    \caption{
    Signatures of the correlated Fermi liquid to the pseudogap doping-driven transition in the local spectral weight, for $U=8.6$ and $\beta=15$.}
    \label{fig:cFL-PG-limit}
\end{figure}

\begin{figure}
    \centering
    \includegraphics[width=0.9\linewidth]{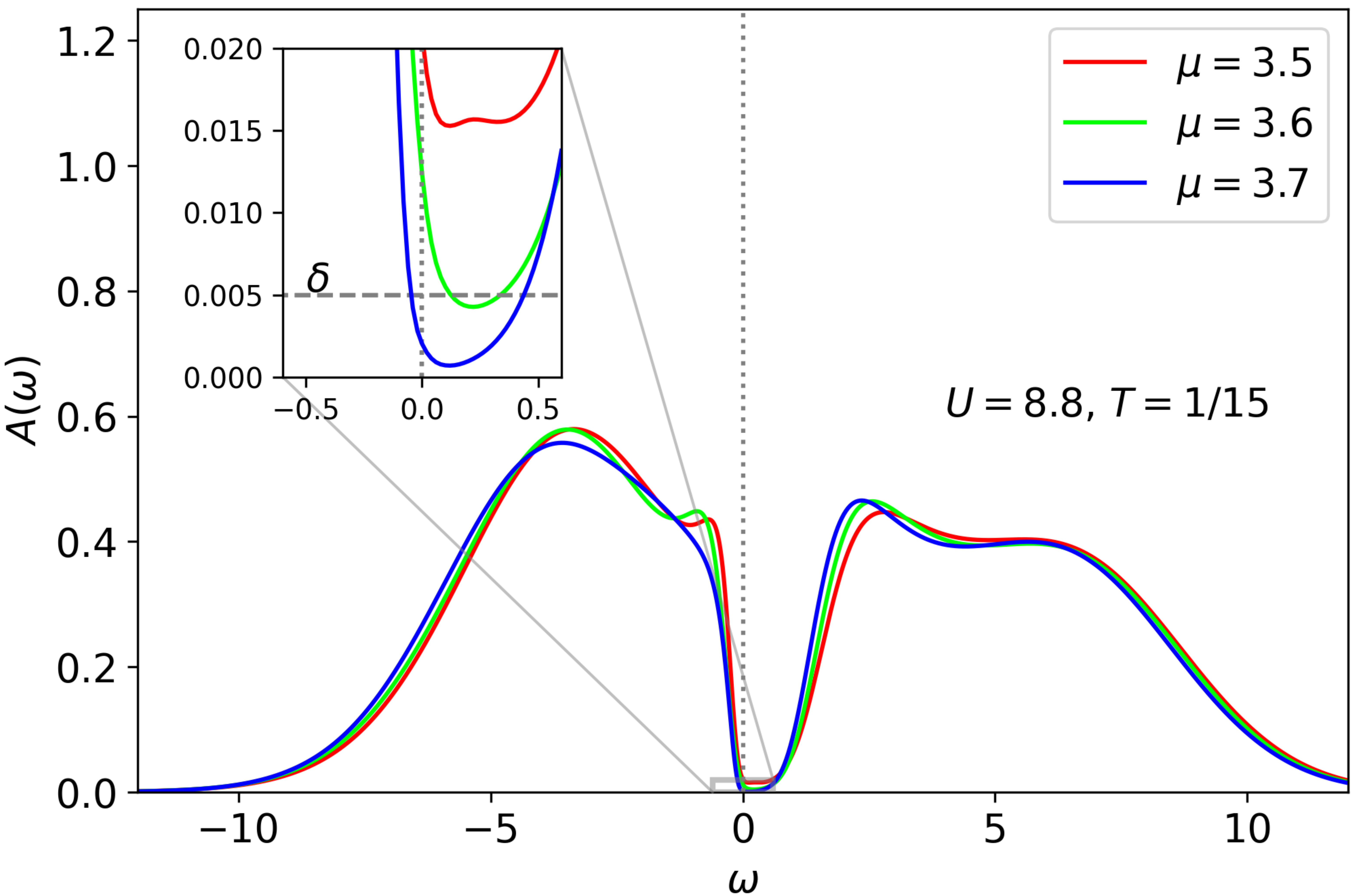}
    \caption{
    Complete gaping of the states at the Fermi level during the doping-driven transition from pseudogap to Mott insulator, at $U=8.8$ and $\beta=15$. The criterion that the residual spectral weight at the Fermi level $A(\omega=0)$ be smaller than $\delta=0.005$ is satisfied between $\mu=3.6$ and $\mu=3.7$, which is equivalent to the red point at $\mu=3.65\pm0.05$ on the phase diagram of Fig.~\ref{fig:mu_U}~(a).}
    \label{fig:Mott-PG-limit}
\end{figure}

% \section{Phase transitions}
\section{PHASE TRANSITIONS}

Defining the limit between the phases found in Fig.~3 of the main text is not always straightforward, especially in this work where we are faced mostly with crossovers. Here, we detail how we defined the transition between all of the phases we explored.

\subsection{Correlated Fermi liquid to pseudogap transition}

We define the correlated Fermi liquid state as a state with a clear peak of local spectral weight near the Fermi level. In a transition towards a pseudogap, only a portion of the states at the Fermi surface get gapped out, thus we expect a minimum in the density of states at the Fermi level, with peaks on each side.
We study this transition by inspecting the evolution of the density of states. We compute the spectral weight by analytically continuing data from the imaginary to the real frequencies using maximum entropy~\cite{MaxEntBergeron}, but the precise doping at which one considers that enough spectral weight has been lost to be considered a pseudogap can we ambiguous when the transition is continuous. Thus, we represent the transition from cFL to PG by a gradient in Fig.~\ref{fig:mu_U}. Figure~\ref{fig:cFL-PG-limit} shows an example of this transition.

\subsection{Pseudogap to the Mott insulator transition}

The transition from the pseudogap to the Mott insulating phase is characterized by a opening of a gap for all states at the Fermi level.
An example of this transition is shown in Fig.~\ref{fig:Mott-PG-limit}. Again, we inspect the local spectral weight obtained by analytical continuation that broadens the edge of a gap and introduces residual spectral weight $\delta$. We decided in this work to define the Mott insulator as satisfying $A(\omega=0) < \delta \equiv 0.005$.
The red dotted line displayed in the phase diagrams of Fig.~\ref{fig:mu_U} are obtained using this criterion. 
Note that changing $\delta$ between $0.002$ and $0.010$ does not impact significantly the position of these lines.

\typeout{}
% \bibliography{references.bib}

\end{document}